# Irradiation effects in the Ni-17Mo-7Cr alloy bombarded with MeV Au ions*

GUO Peng(郭鹏)[1]   YAN Long(闫隆)[2]   HUANG Qing(黄庆)[3]   LI Zhi-Jun(李志军)[2]   HUAI Ping(怀平)[2]   ZHOU Xing-Tai(周兴泰)[2]   XUE Jing-Ming(薛建明)[1;1)]

[1] State Key Laboratory of Nuclear Physics and Technology, Institute of Heavy Ion Physics, Peking University, Beijing 100871, China
[2] Shanghai Institute of Applied Physics, Chinese Academy of Sciences, 2019 Jialuo road, Jiading district, Shanghai 201800, China
[3] Ningbo Institute of Material Technology and Engineering, Chinese Academy of Sciences, Ningbo 315201, Zhejiang, China

**Abstract**

Irradiation effects in Ni-17Mo-7Cr alloy, which is an newly developed structural material for molten salt reactor (MSR), have been systematically investigated by using 3MeV Au ions at different fluences, corresponding to dpa number (displacement per atom) of 1~ 30. GIXRD measurement indicates that the microstrain of the irradiated samples increased from 0.14% to 0.22% as dpa increased from 1 to 30. In the meanwhile, nanoindentation results reveal the Ni-17Mo-7Cr alloy underwent radiation-induced hardening first and then softening at dpa of 30. The swelling rate of Ni-17Mo-7Cr alloy was found around 1.3% at 30 dpa, which means only 0.04% per dpa. Besides, Raman spectra shows that carbon segregation appeared after Au ions irradiation. Our results are very helpful for understanding irradiation damages in Nickel-base alloys, especially for those in purpose of being used in future MSR nuclear energy system.

## 1 Introduction

Nickel-base alloys have excellent high temperature performance and corrosion resistance, and have been playing an important role in advanced nuclear energy systems such as MSR and SCWR [1]. As the structural material in the nuclear energy systems, Nickel-base alloys will be subjected to harsh work conditions especially intensive radiation of neutrons which could be as high as a hundred dpa [2], thus the irradiation resistance of these alloys become one of the key performance for their application in new generation nuclear systems.

Rowcliffe et al. drew some perspectives on radiation effects of a series of Nickel-base alloys for applications in advanced reactors, including alloy718, 706 and PE16, and suggested modifying composition or microstructure of existing alloys or designing completely new alloys by creating a high number density of stable nano-scale clusters or particles to mitigate the effects of irradiation on the performance of the present commercial Nickel-base alloys [3]. Angeliu et al. had systematically studied radiation-induced swelling, creep and embrittlement behavior of Nickel-base alloys such as PE16, Hastelloy X, etc, and recommended that nuclear grade Ni-base alloys be pursued as a method to mitigate radiation-induced embrittlement by reducing helium generation and solute segregation [4]. Hunn et al. studied the irradiation hardening behavior of Fe, He , H ions implanted Inconel 718 prepared by two different heat treatment (solution treatment and aging treatment) by nanoindentation, and found that after Fe ions irradiation, specimens with solution treatment exhibited significant hardening while the specimens with aging treatment exhibited softening as the irradiation dose increased because of the radiation-induced dissolution of $\gamma'$ and $\gamma''$ precipitation [5]. Jin et al .calculated the average grain size and microstrain of C-276 after 82.5dpa $Ar^+$ irradiation by GIXRD and found that the average grain size calculated coincided with the TEM results pretty well and the value of the microstrain was about 0.696% [6].

Currently, the main problems with Ni-base alloys is the radiation embrittlement [7], swelling and phase instability under neutron radiation environment and although a number of studies on the performance of Nickel-base alloys under irradiation of neutron have been carried out [8-12], compared to austenitic stainless steels and ferritic-martensitic steels, irradiation-performance data for Ni-based alloys is very limited. The present work focused on the irradiation effects of a newly developed

*Supported by the the Joint Fund of the National Natural Science Foundation of China and the China Academy of Engineering Physics (Grant No. U1230111), the National Natural Science Foundation of China (Grant No. 91226202).

1) E-mail: jmxue@pku.edu.cn.

Ni-17Mo-7Cr alloy, which is intended to be used in future MSR nuclear energy system. Changes of microstrain, swelling and nanohardness have been measured after samples irradiated with MeV Au ions at different dpa up to 30. These results are very helpful for understanding irradiation damages in nickel-base alloys and for improving irradiation resistance of nickel alloys in the future.

**2 Experimental details**

The Ni-17Mo-7Cr alloy has a nominal chemical composition as shown in Table 1, and the specimens used in experiments were cut into $5\times5\times2$ mm$^3$ and then polished to a mirror-like surface. In the irradiation experiments, part of the specimen surface was masked so that a step should be produced after irradiation, which could be used to calculate sample swelling rate. 3MeV Au$^{2+}$ was used to irradiate the specimens at ion fluences of $8\times10^{13}$cm$^{-2}$, $2.5\times10^{14}$cm$^{-2}$, $7.4\times10^{14}$cm$^{-2}$ and $2.3\times10^{15}$cm$^{-2}$, corresponding to 1dpa, 3dpa, 10dpa and 30dpa, respectively. Ion beam current density was kept in a low level (only around 15nA/cm$^2$) so that the sample temperature did not significantly increase during ion irradiation.

Table1. Chemical composition of the Ni-17Mo-7Cr alloy (wt %)

| element | C | Si | Mn | Fe | Cr | Mo | Ni |
|---|---|---|---|---|---|---|---|
| content | 0.05 | 0.5 | 0.5 | 4 | 7 | 17 | bal |

It is well known that atomic displacement damage caused by energetic ions is not distributed homogenously inside the irradiated sample, we calculated the damage depth profile with SRIM 2006 [13], and the result of $8\times10^{13}$cm$^{-2}$ is shown in Fig. 1. The displacement energy of 40eV was used in the calculation [14-16]. The radiation damage distribution (depth) is up to 600 nm, but within the first 400 nm the damage distribution does not have too much difference and we use the maximum dpa but not the averaged dpa value to stand for the irradiation damage level.

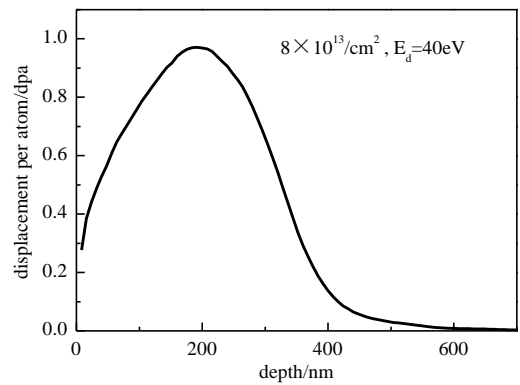

Fig. 1. SRIM 2006 calculation of the damage profile produced by 3MeV Au ion irradiation in Ni-17Mo-7Cr alloy

The structure of the irradiated specimens was determined by grazing incident X-ray diffraction (GIXRD, D8) at a grazing angle of 2° with a Cu K$_\alpha$ radiation source (the wavelength λ=0.15406 nm). The surface morphologies and the swelling behavior were investigated by atomic force microscopy (AFM). Raman Spectrometer (JY HR800) was used to characterize the carbon segregation of the pristine and ion-irradiated Ni-17Mo-7Cr alloy. The excitation source was a 532 nm Ar-ion laser with power 12 mW. The relative effect of the various irradiations on hardening of the implanted layer was measured by G200 nanoindentor in the continuous stiffness mode [17] to measure the hardness as a function of depth. For these tests, the Berkovich diamond indenter tip was used.

**3. Results and discussion**
**3.1 GIXRD analyses of the structure and the microstrain of irradiated samples**

First we investigated the structure changes of the irradiated samples with GIXRD as shown in Fig. 2a. The GIXRD patterns has been normalized for comparison. As dpa increases from 0 to 30, the intensities of all the diffraction peaks decrease but their widths increase. The change of peak intensity indicates that the crystal structure of the irradiated samples has been disrupted by ion irradiation produced defects thus reducing the extent of the ideal crystal structure. More important is that we can obtain the microstrain of irradiated samples by measuring the broadening of diffraction peaks.

The microstrain can be calculated with the following equation [18] :

$$\varepsilon = \frac{\beta(2\theta) - \Delta(2\theta)}{4\tan\theta} \quad , \tag{1}$$

where $\beta(2\theta)$ is the diffraction broadening, θ is the diffraction angle and $\Delta(2\theta)$ is the instrumental broadening. We have calculated the microstrain in the specimens at different dpa, and the results are shown in Fig.2b.

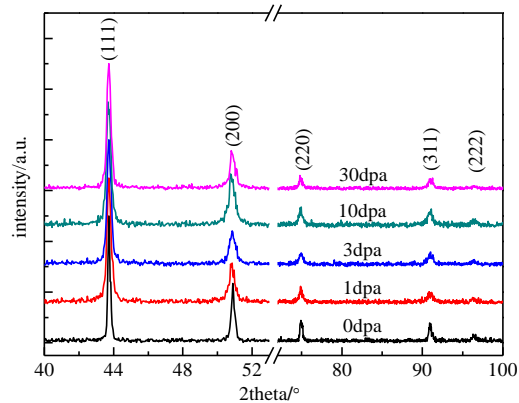

Fig. 2a. (color online)The normalized GIXRD patterns of Ni-17Mo-7Cr alloy specimens irradiated with 3MeV Au ions: from bottom to top, the dpa number is 0,1,3,10,30, respectively

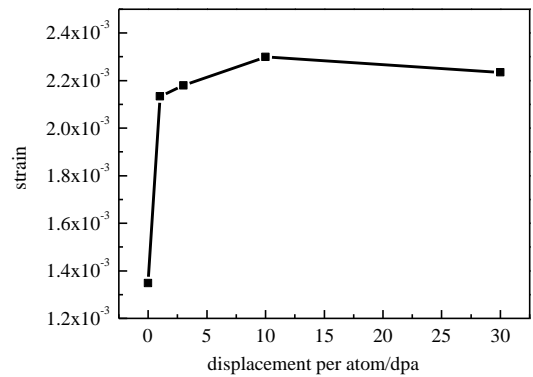

Fig. 2b. Microstrain of Ni-17Mo-7Cr alloy specimens irradiated at different dpa.

Fig. 2b shows that the microstrain of all the irradiated specimens increases significantly compared to the pristine one. The microstrain jumps from 0.14% of the unirradiated sample to 0.21% of the sample irradiated at dpa of 1, then it increases slowly while further increasing ion fluence. It is common sense that the collision cascades caused by implanted ions introduce vacancies, interstitials and substitutional defects. All of these effects can induce a larger lattice strain. When irradiation damage exceeds 1dpa, small vacancy and interstitial clusters (black dot defects) maybe saturate [5,19,20] and there is a considerable number of substitutional defects, thus the implanted ions tend to enter the lattice gap more easily so as to bring little lattice distortion. This tendency was also found in Zr-base alloy irradiated with O ions, in which the microstrain of the specimens increased quickly at lower ion fluences while more slowly when ion fluences were high [21].

**3.2 Swelling rate of irradiated samples**

Swelling is one of the most important effects for judging material's resistance to irradiation damage, because it does not only change sample's shape but also relate to sample's other properties, such as thermal conductivity. We estimated the swelling rate of Ni-17Mo-7Cr alloy by measuring the step height near the interface between the irradiated and masked region using AFM. The results of the sample irradiated at 30dpa are shown in Fig.3.

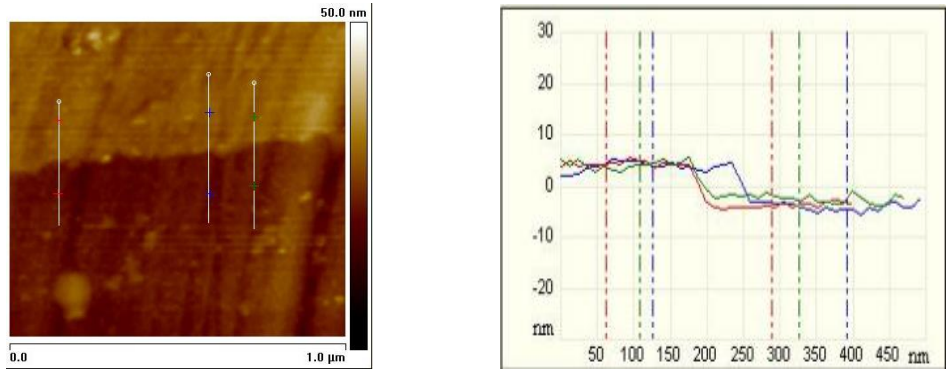

Fig. 3.(color online)(a) The AFM image of the irradiated and unirradiated zone of the Ni-17Mo-7Cr alloy specimen;(b) The measured step height

The averaged step height is 8nm. Considering the thickness of damaged layer is about 600nm, the swelling rate of Ni-17Mo-7Cr alloy is about 1.3% at 30dpa. It is worth mentioning that 30dpa is the value of irradiation damage at the damage peak and the average dpa in the specimen is much less than that. In other words, the swelling rate should be larger than 1.3% if the specimen were irradiated homogenously at 30dpa.

The swelling rate of 1.3% is similar with the values obtained in other kinds of Nickel-base alloys. For example, the swelling rate of PE16 is around 1% when irradiated up to 44dpa while the content of Ni is between40%-50% [9,22], and the swelling rate of alloy 625, Hastelloy X and alloy 800 is <0.6%, 0.6-2.1% and 0.5-5% under 25dpa irradiation respectively [23].

### 3.3 Hardness changes of irradiated samples measured with a nanoindentor

We also measured the nanohardness of irradiated specimens with a nanoindentor. The near surface hardness as a function of the contact depth of the indenter tip is shown in Fig. 4a. A systematic radiation-induced hardening and softening are clear evident.

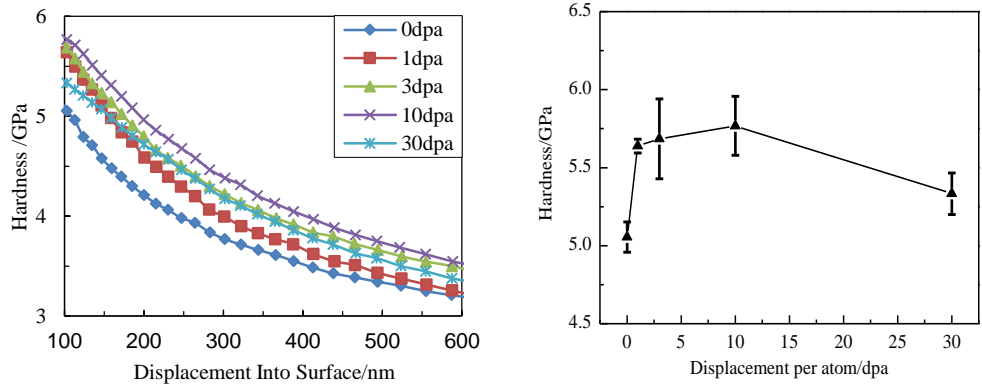

Fig. 4. (color online)(a)The near surface hardness as a function of contact depth, as measured by the nanoindenter; (b)The hardness of the specimens as a function of dpa at 100nm contact depth

The Berkovich diamond indenter tip sampled the hardness in the region extending down about seven times the indenter's contact depth [24] and to eliminate the surface effects and uncertainty in the indenter geometry, a contact depth of 100nm was used in comparing the hardness as a function of irradiation dose as shown in Fig. 4b.

The hardness of the pristine specimen at 100nm contact depth is about 5.06GPa. After irradiation, the hardness increases to 5.64GPa, 5.69GPa and 5.77GPa, corresponding to relative hardening of 11.5%, 12.5% and 14% at 1, 3 and 10dpa, respectively. However, when irradiated to 30dpa, a significant softening can be observed and the hardness drops down to 5.33Gpa, corresponding to relative hardening of 5.4%, lower than all the other three irradiated specimens. The initial hardening may due to the "black dots" (small vacancy and interstitial clusters) defects appears at very low dose (~0.001dpa) [25]. The "black dots" defects could pinned dislocations and inhibit the motion of dislocations. The density of theses "black dots" increases with displacement damage and saturates at about 1dpa. At higher dose, Frank-type loops evolve and continue to produce hardening at a slower rate.

The softening at the highest dose may be caused by the radiation-induced dissolution of precipitation in the specimens [26].

We noted that the change tendency of hardness at different ion fluences is similar with that of microstrain as shown in fig.2b, and both of them reach their maximum at dpa of 10 and then drop a little at dpa of 30. This means a larger microstrain causes a higher hardness, and they are all the results of irradiation defects produced by the incident MeV Au ions.

**3.4 Raman analyses of carbon segregation during ion irradiation**

In order to analyze the radiation-induced segregation of carbon inside the alloy, Raman spectra of the Ni-17Mo-7Cr alloy specimens have been measured, and the results are shown in Fig. 5a.

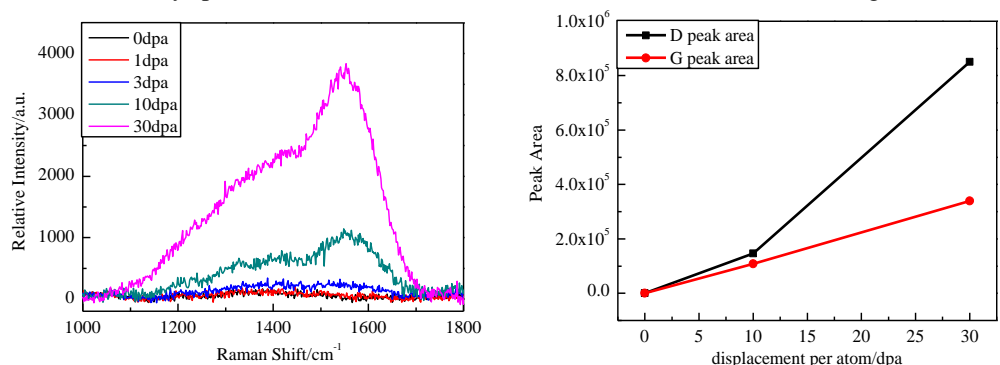

Fig. 5. (color online) (a)Raman spectra of the Ni-17Mo-7Cr alloy specimens irradiated at different ion fluences; (b) Areas of D and G peaks at different dpa.

Raman spectra shows great sensitivity to the detection of carbon atoms. Although the density of carbon atoms in the pristine specimens is 0.5%wt, the carbon atoms are dispersed in the specimens so that there is no C-C vibration peak observed before irradiation. After ion irradiation, peaks near $1355cm^{-1}$ and $1580cm^{-1}$ appear, which correspond to D-peak and G-peak of carbon materials, respectively [27-29]. Carbon peaks in the Raman spectra represent the carbon clusters formed by carbon atoms. We calculated the D peak area and the G peak area of the specimens at 10dpa and 30dpa, and the results are shown in Fig. 5b. It is clear that as the ion fluence increases, the relative Raman intensity of the carbon peaks increases. This is because irradiation damage causes more atomic displacements at higher ion fluence thus carbon atoms have more chance to segregate. The ratio of the integral intensity of the D-peak and G-peak $I_D/I_G$ also increases with ion fluence, which manifests the defects of clustered carbon caused by ion irradiation increased and the degree of graphitization decreased during irradiation.

**4 Conclusions**

After 30dpa $Au^{2+}$ irradiation, the GIXRD result of Ni-17Mo-7Cr alloy specimens indicated that the lattice structure was disrupted, and the radiation-induced microstrain was as low as 0.2%. Small amount of carbon segregation happened in all the specimens and the amount of segregated carbon increased with doses which also happened in other materials [30]. The average swelling rate of Ni-17Mo-7Cr alloy irradiated to 30dpa was about 1.3%, which was acceptable compared to existing data of neutron irradiated Ni-base alloys. In accordance with the results of 316LN steel, the radiation-induced hardening was significant below 1dpa and the relation between the softening behavior of the specimen irradiated to 30dpa and the evolution of microstructure needs further study.

**References**

1  Yvon P and Carré F. J. Nucl. Mater., 2009, **385**: 217
2  Allen T, Busby J, Meyer M et al. Mater. Today, 2010, **13**: 14
3  Rowcliffe A F, Mansur L K, Hoelzer D T et al. J. Nucl. Mater., 2009, **392**: 341
4  Angeliu T M, Ward J T and Witter J K. J. Nucl. Mater., 2007, **366**: 223
5  Hunn J D, Lee E H, Byun T S et al. J. Nucl. Mater., 2001, **296**: 203
6  Jin S X, Guo L P, Yang Z et al. Plasma Sci. Technol., 2012, **14**: 548

7　Murty K and Charit I. J. Nucl. Mater., 2008, **383**: 189
8　Barnaby J, Barton P J, Boothby R M et al. Radiation Effects in Breeder Reactor Structural Materials, 1977, 157.
9　Boothby R. J. Nucl. Mater., 1996, **230**: 148
10　Byun T S and Farrell K. J. Nucl. Mater., 2003, **318**: 292
11　Lindgren J R. Nucl. Technol., 1984, **66**: 607
12　Sencer B H, Bond G M, Garner F A et al. J. Nucl. Mater., 2001, **296**: 145
13　Biersack J P and Ziegler J F. The Stopping and Range of Ions in Solids, 1982, pp.122-156
14　Hashimoto N, Hunn J D, Byun T S et al. J. Nucl. Mater., 2003, **318**: 300
15　Hunn J D, Lee E H, Byun T S et al. J. Nucl. Mater., 2001, **296**: 203
16　Norgett M J, Robinson M T and Torrens I M. Nucl. Eng. Des., 1975, **33**: 50
17　Oliver W C and Pharr G M. J. Mater. Res., 1992, **7**: 1564
18　Williamson G and Hall W. Acta Metall., 1953, **1**: 22
19　Lee E H, Hunn J D, Hashimoto N et al. J. Nucl. Mater., 2000, **278**: 266
20　Lee E H, Hunn J D, Rao G R et al. J. Nucl. Mater., 1999, **271–272**: 385
21　Mukherjee P, Sarkar A and Barat P. Mater. Charact., 2005, **55**: 412
22　Bates J and Powell R. J. Nucl. Mater., 1981, **102**: 200
23　Appleby W K, Sandusky D W and Wolff U E. J. Nucl. Mater., 1972, **43**: 213
24　Samuels L E and Mulhearn T O. J. Mech. Phys. Solids., 1957, **5**: 125
25　Lee E H, Hunn J D, Byun T S et al. J. Nucl. Mater., 2000, **280**: 18
26　Ghetta V, Gorse D and Mazière D. Materials Issues for Generation IV Systems (Netherlands:Springer), 2008, pp. 537-557
27　Baratta G A, Arena M M, Strazzulla G et al. Nucl. Instrum. Methods Phys. Res., Sect. B., 1996, **116**: 195
28　Elman B, Dresselhaus M, Dresselhaus G et al. Physical Review B, 1981, **24**: 1027
29　Mathew S, Joseph B, Sekhar B R et al. Nucl. Instrum. Methods Phys. Res., Sect. B, 2008, **266**: 3241
30　Zhang Y, Qian X, Wang X et al. Nucl. Instrum. Methods Phys. Res., Sect. B, 2013, **297**: 35